\documentclass{article}
\usepackage{arxiv}

\usepackage{url}
\usepackage[hidelinks]{hyperref}
\usepackage[utf8]{inputenc}
\usepackage[small]{caption}
\usepackage{graphicx}
\usepackage{amsmath}
\usepackage{booktabs}
\usepackage{algorithm}
\usepackage{algorithmic}
\usepackage{color}
\usepackage{tikz}

\usepackage{lmodern}

\title{Multidimensional Contrast Limited Adaptive Histogram Equalization}

\author{
    Vincent Stimper$^{1,2,}$\footnote{Contact Authors}\and
    Stefan Bauer$^1$\and
    Ralph Ernstorfer$^3$\and
    \\
    Bernhard Sch{\"o}lkopf$^1$\And
    R. Patrick Xian$^{3,*}$
    \\
    \affiliations
    $^1$Max Planck Institute for Intelligent Systems, 72076 T{\"u}bingen, Germany\\
    $^2$Physics Department, Technical University Munich, 85748 Garching, Germany\\
    $^3$Fritz Haber Institute of the Max Planck Society, 14195 Berlin, Germany\\
    \emails
    \{vincent.stimper,stefan.bauer,bs\}@tue.mpg.de,
    \{ernstorfer,xian\}@fhi-berlin.mpg.de
}

\newcommand\copyrighttext{%
  \footnotesize \textcopyright 2019 IEEE. Personal use of this material is permitted.  Permission from IEEE must be obtained for all other uses, in any current or future media, including reprinting/republishing this material for advertising or promotional purposes, creating new collective works, for resale or redistribution to servers or lists, or reuse of any copyrighted component of this work in other works.}
\newcommand\copyrightnotice{%
\begin{tikzpicture}[remember picture,overlay]
\node[anchor=south,yshift=10pt] at (current page.south) {\fbox{\parbox{\dimexpr\textwidth-\fboxsep-\fboxrule\relax}{\copyrighttext}}};
\end{tikzpicture}%
}

\begin{document}

\maketitle
\copyrightnotice

\begin{abstract}
Contrast enhancement is an important preprocessing technique for improving the performance of downstream tasks in image processing and computer vision. Among the existing approaches based on nonlinear histogram transformations, contrast limited adaptive histogram equalization (CLAHE) is a popular choice for dealing with 2D images obtained in natural and scientific settings. The recent hardware upgrade in data acquisition systems results in significant increase in data complexity, including their sizes and dimensions. Measurements of densely sampled data higher than three dimensions, usually composed of 3D data as a function of external parameters, are becoming commonplace in various applications in the natural sciences and engineering. The initial understanding of these complex multidimensional datasets often requires human intervention through visual examination, which may be hampered by the varying levels of contrast permeating through the dimensions. We show both qualitatively and quantitatively that using our multidimensional extension of CLAHE (MCLAHE) simultaneously on all dimensions of the datasets allows better visualization and discernment of multidimensional image features, as demonstrated using cases from 4D photoemission spectroscopy and fluorescence microscopy. Our implementation of multidimensional CLAHE in Tensorflow is publicly accessible and supports parallelization with multiple CPUs and various other hardware accelerators, including GPUs.
\end{abstract}


\section{Introduction}
\label{sec:introduction}
Contrast is instrumental for visual processing and understanding of the information content within images in various settings \cite{Olshausen2000}. Therefore, computational methods for contrast enhancement (CE) are frequently used to improve the visibility of images \cite{Pratt2007}. Among the existing CE methods, histogram transform-based algorithms are popular due to their computational efficiency. Natural images with a high contrast often contain a balanced intensity histogram, this conception led to the development of histogram equalization (HE) \cite{Hall1974}. A widely adopted example in this class of CE algorithms is the contrast limited adaptive histogram equalization (CLAHE) \cite{Pizer1987,Zuiderveld1994}, originally formulated in 2D, which performs local adjustments of image contrast with low noise amplification. The contrast adjustments are interpolated between the neighboring rectilinear image patches called kernels and the spatial adaptivity in CLAHE is achieved through selection of the kernel size. The intensity range of the kernel histogram (or local histogram) is set by a clip limit that restrains the noise amplification in the outcome. Accounts of the historical development are given in Section II. The use cases of CLAHE and its variants range from underwater exploration \cite{Hitam2013}, breast cancer detection in X-ray mammography \cite{Pisano1998,Dabass2019}, biometric authentication \cite{Yildiz2019}, video forensics \cite{Xiao2019} to charging artifact reduction in electron microscopy \cite{Sim2010} and multichannel fluorescence microscopy \cite{McCollum2007}. Due to the original formulation, its applications are concentrated almost exclusively in fields and instruments producing 2D imagery.

However, the current data acquisition systems are capable of producing densely sampled image data at three or higher dimensions at high rates \cite{Ouyang2017,Pantazis2014,Gao2016,Ersen2015,Schonhense2015}, following the rapid progress in spectroscopic and imaging methods in the characterization of materials and biological systems. Sifting through the image piles to identify relevant features for scientific and engineering applications is becoming an increasingly challenging task. Despite the variety of experimental techniques, the parametric dependence (with respect to time, temperature, pressure, wavelength, concentration, etc.) in the measured system resulting from internal dynamics or external perturbations are often translated into intensity changes registered by the imaging detector circuit \cite{Moore2009}. Visualizing and extracting multidimensional image features from acquired data often begin with human visual examination, which is influenced by the contrast determined by the detection mechanism, specimen condition and instrument resolution. To assist multidimensional image processing and understanding, the existing CE algorithms formulated in 2D should adapt to the demands in higher dimensions (3D and above). Recently, a 3D extension of CLAHE acting simultaneously on all dimensions has been described and shown to compare favorably over 2D CLAHE for volumetric (3D) imaging data both in visual inspection and in a contrast metric, the peak signal-to-noise ratio (PSNR) \cite{Amorim2018}.

The formulations of 2D \cite{Zuiderveld1994} and 3D CLAHE \cite{Amorim2018} algorithms include individual treatment of the image boundaries (corners and the various kinds of edges), which becomes tedious in higher dimensions. In addition, the scalable computation of kernel histograms and intensity transforms presents a major challenge in higher dimensions. In this work, we formulate and implement multidimensional CLAHE (MCLAHE), a flexible and efficient generalization of the CLAHE algorithm to an arbitrary number of dimensions. The MCLAHE algorithm introduces a unified formulation of the image boundaries, allows the use of arbitrary-shape rectilinear kernels and expands the spatial adaptivity of CLAHE to the intensity domain with adaptive histogram range selection. The parallelized implementation of MCLAHE also enables hardware-dependent computational speed-up through the use of multiple CPUs and GPUs. None of these aspects pertaining to handling complex multidimensional imagery have garnered attention in the original formulation of 2D \cite{Pizer1987,Zuiderveld1994} or 3D CLAHE \cite{Amorim2018}. Next, we demonstrate the effectiveness of MCLAHE using visual comparison and computational contrast metrics of two 4D (3D+time) datasets in materials science by photoemission spectroscopy \cite{Hufner2003} and in biological science by fluorescence microscopy \cite{Kubitscheck2017}, respectively. These two techniques are representatives of the current capabilities and complexities of multidimensional data acquisition methods in natural sciences. The use and adoption of CE in their respective communities will potentially benefit visualization and downstream data analysis. Specifically, in the photoemission spectroscopy dataset of electronic dynamics in a semiconductor material, we show that MCLAHE can drastically reduce the intensity anisotropy and enable visual inspection of dynamical features across the bandgap. In the fluorescence microscopy dataset of a developing embryo \cite{Faure2016}, we show that MCLAHE improves the visual discernibility of cellular dynamics from sparse labeling. In addition, we provide a Tensorflow \cite{tensorflow2015} implementation of MCLAHE publicly accessible on GitHub \cite{Stimper2019}, which enables the reuse and facilitates the adoption of the algorithm in a wider community.

The outline of the paper is as follows. In Section II, we highlight the developments in histogram equalization leading up to CLAHE and the use of contrast metrics in outcome evaluation. In Section III, we go into detail on the differentiators of the MCLAHE algorithm from previous lower-dimensional CLAHE algorithms. In Section IV, we describe the use cases of MCLAHE on 4D photoemission spectroscopy and fluorescence microscopy data. In Section V, we comment on the current limitations and potential improvements in the algorithm design and the software implementation. Finally, in Section VI, we draw the conclusions.

\section{Related work}

Histogram transform-based CE began with the histogram equalization algorithm developed by Hall in 1974 \cite{Hall1974}, where a pixelwise intensity mapping derived from the normalized cumulative distribution function (CDF) of the entire image's intensity histogram is used to reshape the histogram into a more uniform distribution \cite{Hall1974,Sinha2012}. However, Hall's approach calculates the intensity histogram globally, which can overlook fine-scale image features of varying contrast. Subsequent modifications to HE introduced independently by Ketcham \cite{Ketcham1976} and Hummel \cite{Hummel1977}, named the adaptive histogram equalization (AHE) \cite{Pizer1987}, addressed this issue by using the intensity histogram of a rectangular window, called the \textit{kernel}, or the contextual region, around each pixel to estimate the intensity mapping. However, AHE comes with significant computational overhead because the kernel histograms around all pixels are calculated. In performance, the noise in regions with relatively uniform intensities tends to be overamplified. Pizer \textit{et al.} proposed a version of AHE \cite{Pizer1987} with much less computational cost by using only the adjoining kernels that divide up the image for local histogram computation. The transformed intensities are then bilinearly interpolated to other pixels not centered on a kernel. Moreover, they introduced the histogram clip limit to constrain the intensity redistribution and suppress noise amplification \cite{Pizer1987,Zuiderveld1994}. The 3D extension of CLAHE was recently introduced by Amorim \textit{et al.} \cite{Amorim2018} for processing medical images. Their algorithm uses volumetric kernels to compute the local histograms and trilinear interpolation to derive the voxelwise intensity mappings from nearest-neighbor kernels. Qualitative results were demonstrated on magnetic resonance imaging data, showing that the volumetric CLAHE leads to a better contrast than applying 2D CLAHE separately to every slice of the data.

Evaluating the outcome of contrast enhancement requires quantitative metrics of image contrast, which are rarely used in the early demonstrations of HE algorithms \cite{Pizer1987,Hall1974,Ketcham1976,Hummel1977,Dale-Jones1993,Stark2000} because the use cases are predominantly in 2D and the improvements of image quality are largely intuitive. In domain-specific settings involving higher-dimensional (3D and above) imagery, intuition becomes less suitable for making judgments, but computational contrast metrics can provide guidance for evaluation in combination with user objectives. The commonly known contrast metrics include the mean squared error (MSE) or the related PSNR \cite{Wang2004,Campos2019}, the standard deviation (also called the root-mean-square contrast) \cite{Peli1990} and the Shannon entropy (also called the grey-level entropy) \cite{Gu2016}. These metrics are naturally generalizable to imagery in arbitrary dimensions \cite{Kriete1994} and are easy to compute. We also note that despite the recently developed 2D image quality assessment scores based on the current understanding of human visual systems \cite{Wang2004,Gu2016} proved to be more effective than the classic metrics we choose to quantify contrast, their generalization and relevance to the evaluation of higher-dimensional images obtained in natural sciences and engineering settings, often without undistorted references, are not yet explored, so they are not used here for comparison of results.

\section{Methods}

\subsection{Overview}
Extending CLAHE to arbitrary dimensions requires to address some of the existing limitations of the 2D (or 3D) version of the algorithm. (1) The formulations of 2D \cite{Pizer1987,Zuiderveld1994} and 3D CLAHE \cite{Amorim2018} involve explicit enumerated treatment of image boundaries, which becomes tedious and unscalable in arbitrary dimensions because the number of distinct boundaries scales exponentially as $3^D-1$ with respect to the number of dimensions $D$. We resolve this issue by introducing data padding in MCLAHE as an initial step such that every $D$-dimensional pixel has a neighborhood of the same size in the augmented data (see Section III.B). The data padding also enables the choice of kernels with an arbitrary size smaller than the original data. (2) The formalism for calculating and interpolating the intensity mapping needs to be generalized to arbitrary dimensions. We present a unified formulation using the Lagrange form of multilinear interpolation \cite{Phillips2003} that includes the respective use of bilinear and trilinear interpolation in 2D \cite{Pizer1987,Zuiderveld1994} and 3D \cite{Amorim2018} versions of CLAHE as special cases in lower dimensions (see Section III.C). (3) To further suppress noise amplification in processing image data containing vastly different intensity features, we introduce adaptive histogram range (AHR), which extends the spatial adaptivity of the original CLAHE algorithm to the intensity domain. AHR allows the choice of local histogram range according to the intensity range of each kernel instead of using a global histogram range (GHR) (see Section III.D).

\begin{figure}[htb!]
    \centering
    \includegraphics[width=\linewidth]{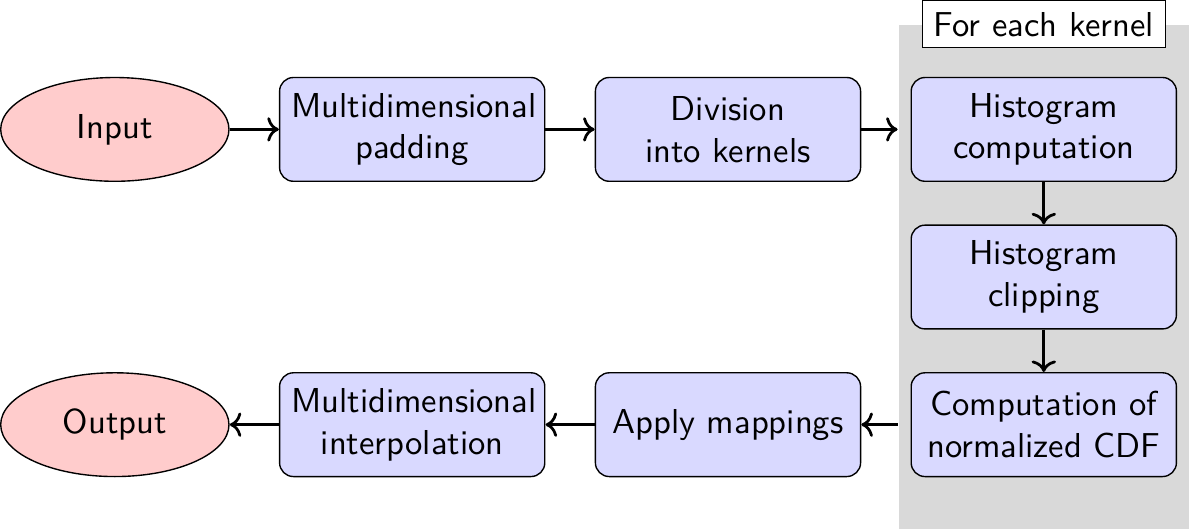}
    \caption{Schematic of the MCLAHE algorithm.}
    \label{fig:schematic}
\end{figure}

The MCLAHE algorithm is summarized graphically in Fig. \ref{fig:schematic} and in pseudocode in Algorithm \ref{alg:mclahe}. It operates on input data of dimension $D$, where $D$ is a positive integer. Let $s_i$ be the size of data along the $i$th dimension, so $i\in\{0,...,D-1\}$. The algorithm begins with padding of the input data around the $D$-dimensional edges. The padded data are then kernelized, or divided into adjoining rectilinear kernels with dimension $D$ and a size of $b_i$ along the $i$th dimension defined by the user. Next, in each kernel, we separately compute and clip its intensity histogram and obtain the normalized CDF. The intensity mapping at each $D$-dimensional pixel is computed by multilinear interpolation of the transformed intensities among the normalized CDFs in the pixel's nearest-neighbor kernels. Finally, the contrast-enhanced output data are generated by applying the intensity transform to every pixel in the input data.

\begin{algorithm}
\caption{Formulation of the MCLAHE algorithm in pseudocode. Here, // denotes the integer division operator, cdf the cumulative distribution function, and map the intensity mappings applied to the high dimensional pixels.}
\label{alg:mclahe}
\textbf{Input}: data\_in\\
\textbf{Parameters}: kernel\_size (array of integers for all kernel dimensions), clip\_limit (threshold value in $\left[0, 1\right]$ for clipping the local histograms), n\_bins (number of bins of the local histograms) \\
\textbf{Output}: data\_out
\begin{algorithmic}[1]
\STATE pad\_len = 2 $\cdot$ kernel\_size - 1 + ((shape(data\_in) - 1) mod kernel\_size)
\STATE data\_hist = symmetric\_padding(data\_in, [pad\_len // 2, (pad\_len + 1) // 2])
\STATE b\_list = split data\_hist into kernels of size kernel\_size
\FORALL{b in b\_list}
\STATE h = histogram(b, n\_bins)
\STATE Redistribute weight in h above clip\_limit equally across h
\STATE cdf\_b = cdf(h)
\STATE map[b] = (cdf\_b - min(cdf\_b)) / (max(cdf\_b) - min(cdf\_b))
\ENDFOR
\FORALL{neighboring kernel}
\FORALL{pixel p in data\_in}
\STATE u = map[b in neighboring kernel of p](p)
\FOR{d = 0 ... D-1}
\STATE u = u $\cdot$ (coefficient of the neighboring kernel in dimension d)
\ENDFOR
\STATE Assign u to pixel p in data\_out
\ENDFOR
\ENDFOR
\STATE \textbf{return} data\_out
\end{algorithmic}
\end{algorithm}

\subsection{Multidimensional padding}
\label{sec:padding}

\begin{figure}[htb!]
    \centering
    \includegraphics[width=\linewidth]{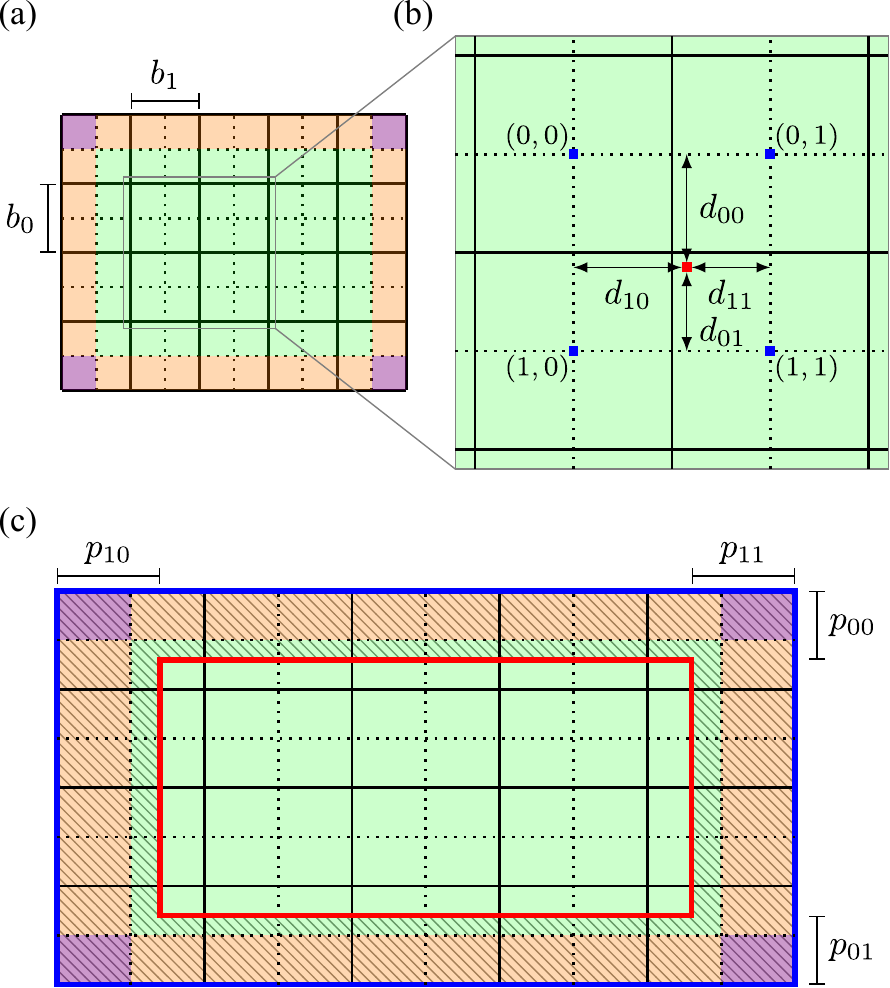}
    \caption{Illustration of the concepts related to the MCLAHE algorithm in 2D. In (a)-(c) the image is equipartitioned into kernels of size $(b_0, b_1)$ bounded by solid black lines. The dotted black lines indicate regions with pixels having the same nearest-neighbor kernels. Color coding is used to specify the types of border regions, with the areas in green, orange and magenta having four, three and two nearest-neighbor kernels, respectively. (a) The original image data in 2D that are divided into kernels. (b) A zoomed-in region of (a). The red square mark in (b) represents a pixel under consideration and the four blue square marks represent the closest kernel centers next to the red one. The distances between the red pixel and the nearest-neighbor kernel centers are labeled as $d_{00}$, $d_{01}$, $d_{10}$, $d_{11}$, respectively. (c) The padded image with the original image now bounded by solid red lines and the padding indicated by the hatchings. The padding lengths in 2D are labeled as $p_{00}$, $p_{01}$, $p_{10}$, $p_{11}$ in (c) and their values are calculated using Eq. \eqref{equ:padding_lengths}.}
    \label{fig:case2d}
\end{figure}

Because of the exponential scaling of the distinct boundaries as $3^D-1$ with respect to the data dimensionality $D$, we use multidimensional padding to circumvent the enumerated treatment of boundaries and ensure that the data can be divided into integer multiples of the user-defined kernel size. The padding is composed of two parts. We discuss the case for $D$ dimensions and illustrate with an example for $D=2$ in Fig. \ref{fig:case2d}. Firstly, we require that the intensity histogram of each kernel is computed with the same number of $D$-dimensional pixels. Therefore, the size of the padded data should be an integer multiple of the kernel size. For each dimension, if $s_i$ is not an integer multiple of $b_i$, a padding of \mbox{$b_i - (s_i \mod b_i)$} is needed. To absorb the case when \mbox{$s_i \mod b_i \equiv 0$}, we add a shift of $-1$ to the expression. Therefore, the padding required along the $i$th dimension of the kernel to make the data size divisible by the kernel size is \mbox{$b_i - 1 - ((s_i - 1) \mod b_i)$}. Secondly, we require that every $D$-dimensional pixel in the original data has the same number of nearest-neighbor kernels such that the pixels at the border do not need a special treatment in the interpolation step. Therefore, we need to pad, in addition, by the kernel size, $b_i$, along the $i$th dimension. To satisfy both requirements, the total padding length along the $i$th dimension, $p_i$, is
\begin{equation}
    p_i = 2b_i - 1 - ((s_i - 1)\mod b_i).
\end{equation}
This length is split into two parts, $p_{i0}$ and $p_{i1}$, and attached to the start and end of each dimension, respectively.
\begin{equation}
    \begin{split}
        p_{i0} &= p_i // 2, \\
        p_{i1} &= (p_i + 1) // 2.
    \end{split}
    \label{equ:padding_lengths}
\end{equation}
Here, the $//$ sign denotes integer division. To keep the local intensity distribution at the border of the image unchanged, the padding is implemented by mirroring the intensities along the boundaries of the data (symmetric padding). The padding procedure is described in lines 1--2 in Algorithm \ref{alg:mclahe}.

\subsection{Multidimensional interpolation}
\label{sec:interpolation}

To derive a generic expression for the intensity mapping in arbitrary dimensions, we start with the example in 2D CLAHE, where each pixel intensity $I_n$ ($n$ being the pixel index) is transformed by a bilinear interpolation of the mapped intensities obtained from the normalized CDF of the nearest-neighbor kernels \cite{Pizer1987,Zuiderveld1994}. We introduce the kernel index $\mathbf{i} = (i_0,i_1)\in\{0,1\}^2$. The values of $0$ and $1$ in the binary alphabet $\{0,1\}$ represent the two sides (i.e. above and below), respectively, in a dimension divided by the pixel in consideration. For the 2D case, the index $(i_0, i_1)$ can take any value of $(0, 0)$, $(0, 1)$, $(1, 0)$ and $(1, 1)$, as shown in Fig. \ref{fig:case2d}(b). Let $m_\mathbf{i}$ be the intensity mapping obtained from the kernel with the index $\mathbf{i}$, then
\begin{equation}
  m_\mathbf{i}(I_n) = \widehat{\text{CDF}}_\mathbf{i}(I_n),
\end{equation}
where $\widehat{\text{CDF}}_\mathbf{i}$ represents the normalized CDF obtained from the clipped histogram of the kernel with the index $\mathbf{i}$. As shown in Fig. \ref{fig:case2d}(b), the bilinear interpolations for the pixel in consideration located at the red square mark are computed using the four nearest-neighbor kernels centered on the blue square mark. The interpolation coefficients, $c_{\mathbf{i}}$, are represented as Lagrange polynomials \cite{Kreyszig2011,Phillips2003} using the kernel size $(b_0, b_1)$ and the distances $(d_{00}, d_{01}, d_{10}, d_{11})$ between the pixel and the kernel centers in the two dimensions.
\begin{align}
    c_{00} &= \frac{(b_0 - d_{00})(b_1 - d_{10})}{b_0 b_1},
    \label{eq:c00}\\
    c_{01} &= \frac{(b_0 - d_{00})(b_1 - d_{11})}{b_0 b_1},
    \label{eq:c01}\\
    c_{10} &= \frac{(b_0 - d_{01})(b_1 - d_{10})}{b_0 b_1},
    \label{eq:c10}\\
    c_{11} &= \frac{(b_0 - d_{01})(b_1 - d_{11})}{b_0 b_1}.
    \label{eq:c11}
\end{align}
The transformed intensity $\Tilde{I}_n$ from $I_n$ is given by,
\begin{align}
    \Tilde{I}_n =& c_{00}m_{00}(I_n) + c_{01}m_{01}(I_n) + c_{10}m_{10}(I_n) \nonumber\\
    &+ c_{11}m_{11}(I_n).
    \label{equ:bilinear}
\end{align}
Eq. \eqref{eq:c00}-\eqref{eq:c11} and \eqref{equ:bilinear} can be rewritten in a compact form using the kernel index $\mathbf{i}$ introduced earlier,
\begin{align}
    \Tilde{I}_n &= \sum_{\mathbf{i}\in\{0,1\}^2}c_\mathbf{i}m_\mathbf{i}(I_n),\\
    c_\mathbf{i} &= \prod_{j=0}^{1}\frac{b_j-d_{j i_j}}{b_j}.\label{equ:coef_2d}
\end{align}
In the 2D case, the term $d_{j i_j}$ takes on the value $d_{j i_0}$ or $d_{j i_1}$. The choice of $i_0$ and $i_1$ from the binary alphabet $\{0,1\}$ in $d_{j i_j}$ follows that of the kernel index $\mathbf{i}$. The special cases of transforming the border pixels in the image are naturally resolved in our case after data padding (see Section III.B).

In $D$ dimensions, the kernel index $\mathbf{i} = (i_0,i_1,...,i_{D-1}) \in \{0,1\}^D$. Analogous to the two-dimensional case described before, the intensity mapping of each $D$-dimensional pixel is now calculated by multilinear interpolation between the $2^D$ nearest-neighbor kernels in all dimensions, mathematically,
\begin{equation}
\begin{split}
    \Tilde{I}_n =& c_{00...0}m_{00...0}(I_n) + c_{00...1}m_{00...1}(I_n) \\
    &+...+c_{11...0}m_{11...0}(I_n) + c_{11...1}m_{11...1}(I_n).
\end{split}
\label{equ:multilinear}
\end{equation}
Similarly, Eq. \eqref{equ:multilinear} and the corresponding expressions for the interpolation coefficients can be written in a compact form using the kernel index $\mathbf{i}$ as
\begin{align}
    \Tilde{I}_n &= \sum_{\mathbf{i}\in\{0,1\}^D}c_\mathbf{i}m_\mathbf{i}(I_n),\\
    c_\mathbf{i} &= \prod_{j=0}^{D-1}\frac{b_j-d_{j i_j}}{b_j}.
\end{align}
The formalism introduced for the 2D case in Eq. \eqref{equ:bilinear}-\eqref{equ:coef_2d} generalizes to arbitrary dimensions with only an update to the kernel index $\mathbf{i}$. In Algorithm \ref{alg:mclahe}, the calculation of intensity mappings through interpolation is described in lines 12--15.

\subsection{Adaptive histogram range}

In the original formulation of CLAHE in 2D \cite{Pizer1987,Zuiderveld1994}, the local histogram ranges for all kernels are the same, which works well when the kernels contain intensities in a similarly wide range. Tuning of the trade-off between noise amplification and the signal enhancement is then achieved through selection of the kernel size and the clip limit \cite{Campos2019}. However, if different patches of the image data contain local features within vastly different but narrow intensity ranges, they may accumulate in very few histogram bins with values specified globally. Accounting for the disparity in CLAHE will require a high clip limit to enhance and therefore comes with the price of noise amplification in many parts of the data. This problem may be ameliorated by adaptively choosing the local histogram range to lie within the minimum and maximum of the intensity values of the kernel, while keeping the number of bins the same for all kernels. An example use case of the AHR is presented in Section IV.A.

\section{Applications}
We now apply the MCLAHE algorithm to two cases in the natural sciences that involve large densely sampled 4D (3D+time) data. Each example includes a brief introduction of the background knowledge on the type of measurement, the resulting image data features and the motivation for the use of contrast enhancement, followed by discussion and comparison of the outcome using MCLAHE. The performance details are provided at the end of each example.

\subsection{Photoemission spectroscopy}
\label{sec:mpes}
\noindent\textbf{Background information.} In photoemission spectroscopy, the detector registers electrons liberated by intense vacuum UV or X-ray pulses from a solid state material sample \cite{Hufner2003}. The measurement is carried out in the so-called 3D momentum space, spanned by the coordinates $(k_x, k_y, E)$, in which $k_x$, $k_y$ are the electron momenta and $E$ the energy. The detected electrons form patterns carrying information about the anisotropic electronic density distribution within the material. The fourth dimension in time-resolved photoemission spectroscopy represent the waiting time in observation by photoemission since the electronic system is subject to an external perturbation (i.e. light excitation). The negative time frames represent the observations taking place before the light excitation. In the image data acquired in photoemission spectroscopy, the inhomogeneous intensity modulation from the experimental geometry, light-matter interaction \cite{Moser2017} and scattering background creates contrast variations within and between the so-called energy bands, which manifest themselves as intercrossing curves (in 2D) or surfaces (in 3D) blurred by convolution with the instrument response function and further affected by other factors such as the sample quality and the dimensionality of the electronic system, etc \cite{Hufner2003}. Visualization and demarcation of the band-like image features are of great importance for understanding the momentum-space electronic distribution and dynamics in multidimensional photoemission spectroscopy \cite{Schonhense2015}. However, in addition to the physical limitations on the contrast inhomogeneities listed before, the intensity difference between the lower bands (or valence bands) and the upper bands (or conduction bands) on the energy scale is on the order of 100 or higher and varies by the materials under study and light excitation conditions. To improve the image contrast in multiple dimensions, we applied MCLAHE to a 4D (3D+time) dataset measured for the time- and momentum-resolved electronic dynamics of tungsten diselenide (WSe$_2$), a semiconducting material with highly dispersing electronic bands \cite{Riley2014}. The 4D data were obtained from an existing experimental setup \cite{Puppin2018} and processed using a custom pipeline \cite{Xian2019a,Xian2019b} from detected single photoelectron events. For comparison of contrast enhancement, we applied both 3D and 4D CLAHE to the 4D photoemission spectroscopy data. In the case of 3D CLAHE, the algorithm was applied to the 3D data at each time frame separately.
\begin{figure*}[htbp!]
    \centering
    \includegraphics[width=0.9\textwidth]{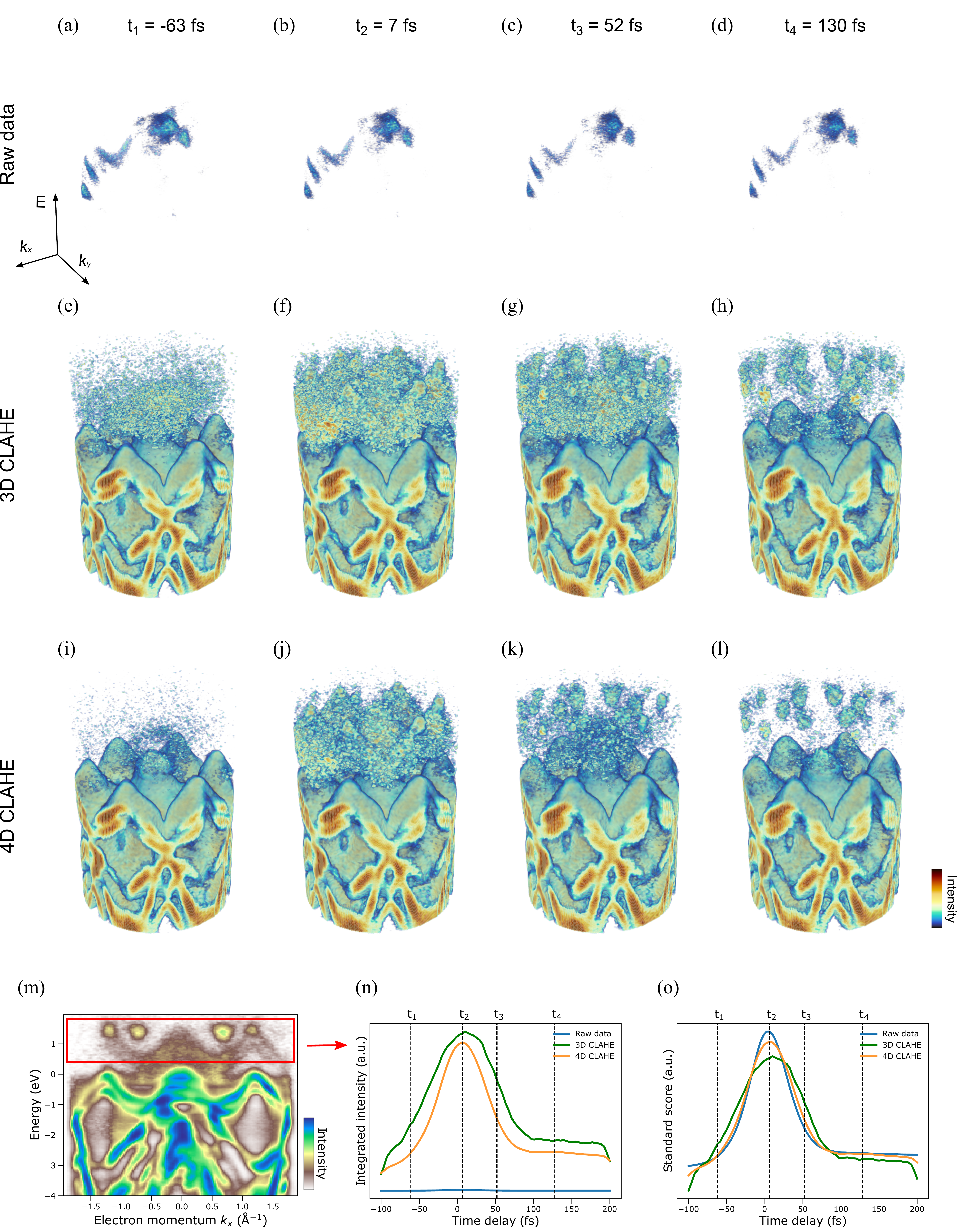}
    \caption{Applications of MCLAHE to 4D (3D+time) photoemission spectroscopy data featuring the temporal evolution of the electronic band structure of the semiconductor WSe$_2$ during and after optical excitation (see Section IV.A). Four time steps in the 4D time series are selected for visualization, including the raw data in (a)-(d), the 3D CLAHE-processed data in (e)-(h) and the 4D CLAHE-processed data in (i)-(l). The adaptive histogram range (AHR) setting in the MCLAHE algorithm were included in the data processing. All 3D-rendered images in (a)-(l) share the same color scaling shown in (l). The integrated dynamics in (n)-(o) over the region specified by the box in (m) over the 3D momentum space show that the 4D CLAHE amplifies less noise while better preserves the dynamical timescale than the 3D CLAHE, in comparison with the original data.}
    \label{fig:photoemission}
\end{figure*}

\noindent\textbf{Results and discussion.} Stills from the raw data and the results are compared in Fig. \ref{fig:photoemission}, along with the contrast metrics computed and listed in Table \ref{tab:metrics_pes}. The supplementary files include videos 1-3 and video 4 for comparing unprocessed and processed data rendered in 2D slices and in 3D, respectively. As shown in Fig. \ref{fig:photoemission}(a)-(d), the original photoemission spectroscopy data are visualized poorly on an energy scale covering both the valence (lower) and conduction (upper) bands. The situation is much improved in the MCLAHE-processed data with AHR setting shown in Fig. \ref{fig:photoemission}(e)-(l), where the population dynamics in the conduction band of WSe$_2$ \cite{Bertoni2016,Puppin2017} and the broadening of the valence bands are sufficiently visible to be placed on the same colorscale, allowing to identify and correlate fine features of the momentum-space dynamics. The improvement in contrast is also reflected quantitatively in Table \ref{tab:metrics_pes} in the drastic changes in standard deviation \cite{Peli1990} between the unadjusted (smoothed) and processed data. On the other hand, the GHR setting of MCLAHE cannot visualize the upper bands well (see comparisons in supplementary videos 1-3) because the regions in the lower and upper bands contain drastically different intensity features. Next, we compare performance between 3D and 4D CLAHE under the AHR setting. The decrease in MSE (0.1121 $\rightarrow$ 0.1050) or, equivalently, the increase in PSNR (147.98 $\rightarrow$ 148.26) shown in Table \ref{tab:metrics_pes} indicates that 4D CLAHE is more suited here because a smaller MSE implies closer resemblance to the original data \cite{Campos2019}. Furthermore, visual inspection of the results in Fig. \ref{fig:photoemission}(e)-(l) and in supplementary videos 1-4 finds less severe noise enhancement when applying 4D CLAHE to the whole dataset than 3D CLAHE to each time frame.
\begin{table}[htb!]
\caption{Contrast metrics for photoemission spectroscopy data}
\setlength{\tabcolsep}{3pt}
\centering
\begin{tabular}{|p{80pt}|p{30pt}|p{30pt}|p{30pt}|p{25pt}|}
\hline
Dataset & MSE & PSNR & STD & ENT \\
\hline
Raw & - & - & 0.0667 & 2.294 \\
Smoothed & 0.0015 & 166.63 & 0.0938 & 2.615 \\
3D CLAHE (GHR) & 0.0426 & 152.18 & 0.2296 & 3.048 \\
4D CLAHE (GHR) & 0.0428 & 152.16 & 0.2299 & 3.049 \\
3D CLAHE (AHR) & 0.1121 & 147.98 & 0.2838 & 4.471 \\
4D CLAHE (AHR) & 0.1050 & 148.26 & 0.2818 & 4.387 \\
\hline
\end{tabular}
\caption*{\\
MSE: mean square error. PSNR: peak signal-to-noise ratio.\\
STD: standard deviation. ENT: Shannon entropy.
}
\label{tab:metrics_pes}
\end{table}

To quantify the influence of contrast enhancement on the dynamical features in the data, we calculated the integrated intensity in the conduction band of the data in all three cases and the results are summarized in Fig. \ref{fig:photoemission}(m)-(o). The standard score in Fig. \ref{fig:photoemission}(o) is used to compare the integrated signals in a scale-independent fashion. The dynamics represented in the intensity changes are better preserved in 4D than 3D CLAHE-treated data and the former are less influenced by the boundary artifacts in the beginning and at the end of the time delay range. The artefactual delays created by 3D CLAHE in the onset and recovery of changes, around $t_1$ and $t_3$ in Fig. \ref{fig:photoemission}(o), respectively, are shown even clearer in the supplementary videos 1-4. These observations reinforce the argument that 4D CLAHE is superior to its 3D counterpart overall in content-preserving contrast enhancement.

\noindent\textbf{Processing details.} The raw 4D photoemission spectroscopy data have a size of 180$\times$180$\times$300$\times$80 in the ($k_x$, $k_y$, $E$, $t$) dimensions. They were first denoised using a Gaussian filter with standard deviations of 0.7, 0.9 and 1.3 along the momenta, energy and time dimensions, respectively. In the applications of both 3D and 4D CLAHE, we set a clip limit of 0.02 and assigned 256 grey-level bins to the local histograms. The kernel size for 4D CLAHE was (30, 30, 15, 20) and for 3D CLAHE the same kernel size for the first three dimensions, or ($k_x$, $k_y$, $E$), was used. Both GHR and AHR settings were tested for comparison. The processing ran on a server with 64 Intel Xeon CPUs at 2.3 GHz and 254GB RAM. The total runtime, including memory copy operations, for processing the whole dataset with 4D CLAHE was about 5.3 mins. In addition, we benchmarked the performance of 4D CLAHE on the GPU (NVIDIA GeForce GTX 1070, 8GB RAM) of the server using the first 25 time frames of the dataset. The total runtime was 34 s with the GPU versus 104 s without the GPU, representing a 3.1-fold speed-up.

\subsection{Fluorescence microscopy}
\noindent\textbf{Background information.} In 4D fluorescence microscopy, the measurements are carried out in the Cartesian coordinates of the laboratory frame, or $(x, y, z)$, with the fourth dimension representing the observation time $t$ since fertilization. In practice, the photophysics of the fluorophores \cite{Valeur2013}, the autofluorescence background \cite{Kubitscheck2017} from the labeled and unlabeled parts of the sample and the detection method, such as the attenuation effect from scanning measurements at different depths or nonuniform illumination of fluorophores \cite{Waters2009}, pose limits on the achievable image contrast in the experiment. The image features in fluorescence microscopy data often include sparsely labeled cells and cellular components such as the nuclei, membranes, dendritic structures, and other organelles. The limited contrast may render the downstream data annotation tasks, such as segmentation, tracking and lineage tracing \cite{Nketia2017,Kretzschmar2012}, challenging. Therefore, a digital contrast enhancement method is potentially useful to improve the visibility of the cells and their corresponding dynamics. We demonstrate the use of MCLAHE for this purpose on a publicly available 4D (3D+time) fluorescence microscopy dataset \cite{Faure2016} of the embryo development of ascidian (\textit{Phallusia mammillata}), or sea squirt. The organism is stained and imaged \textit{in toto} to reveal its development from a gastrula to tailbud formation with cellular resolution \cite{Faure2016}. During embryo development, the fluorescence contrast exhibits time dependence due to cellular processes such as division and differentiation. We use the data from one fluorescence label channel containing the nuclei and process through the MCLAHE pipeline.
\begin{figure*}[htb!]
    \centering
    \includegraphics[width=0.9\textwidth]{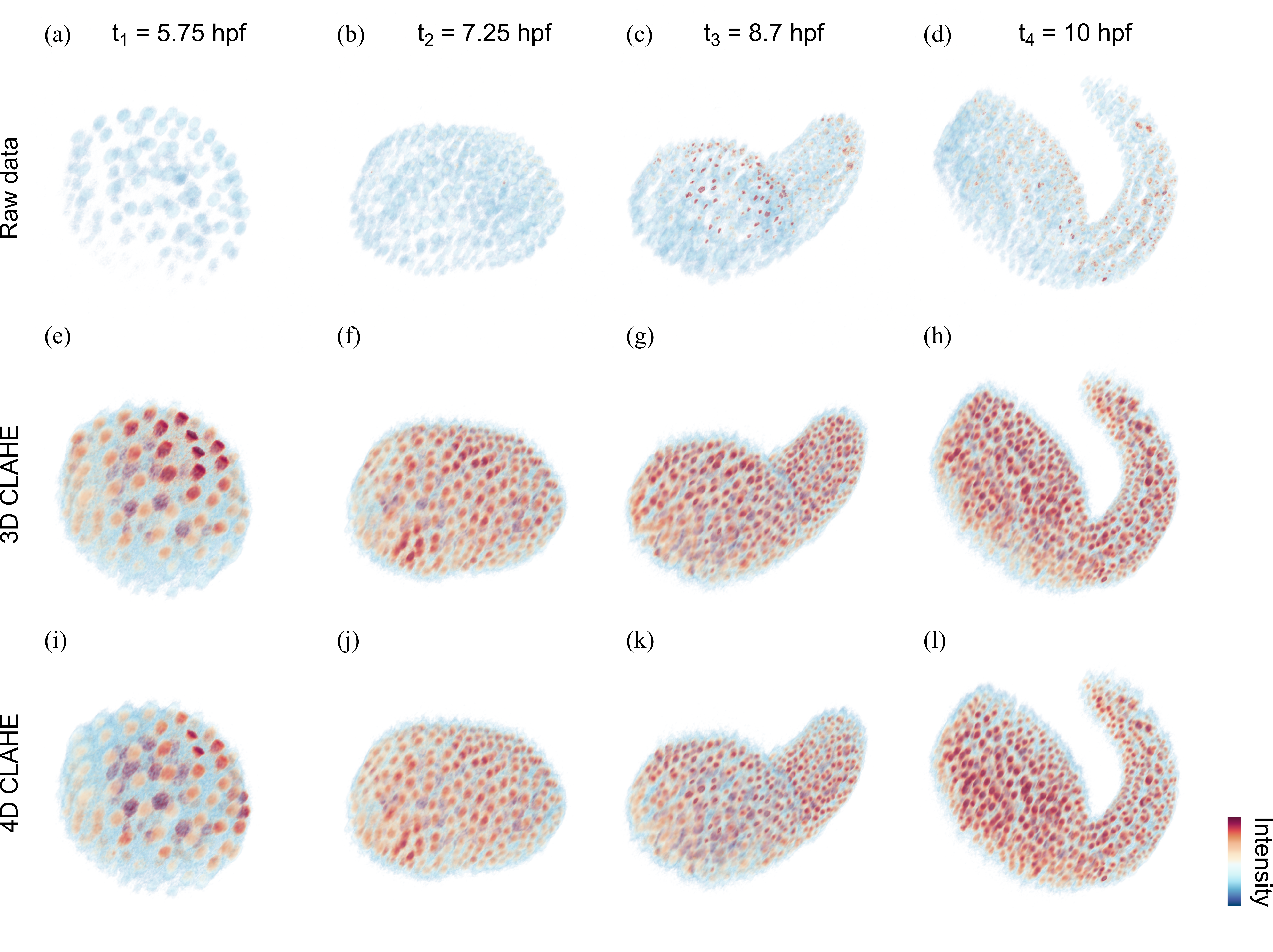}
    \caption{Applications of MCLAHE to 4D (3D+time) fluorescence microscopy data of the embryo development of ascidian (\textit{Phallusia mammillata}), or sea squirt. Four time frames (hpf = hours post fertilization) in the 4D time series are visualized here for comparison, including the raw data in (a)-(d), the 3D CLAHE-processed data in (e)-(h) and the 4D CLAHE-processed data in (i)-(l). All images in (a)-(l) are rendered in 3D with the same axis, orientation and the same color scaling as in (l). Both 3D and 4D CLAHE-processed data show drastic improvement in the image contrast, while the results from 4D CLAHE better preserves the dynamical intensity features from the cellular processes.}
    \label{fig:embryo}
\end{figure*}

\noindent\textbf{Results and discussion.} The results are compared with the original data on the same colorscale in Fig. \ref{fig:embryo} and the corresponding contrast metrics are shown in Table \ref{tab:metrics_fm}. The supplementary files include video 5 and video 6 for comparing the unprocessed and processed data in a 2D slice and in 3D, respectively. As shown in Fig. \ref{fig:embryo}(a)-(d), the intensities in the raw fluorescence microscopy data are distributed very unevenly in the colorscale. The MCLAHE-processed 4D time series show significant improvement in the visibility of the cells against the background signal (e.g. autofluorescence, detector dark counts, etc). This is reflected in the surge in contrast represented by standard deviation as shown in Table \ref{tab:metrics_fm}. In contrast to the previous example, the AHR option in MCLAHE was not used in processing the embryo development dataset because the cellular feature sizes and their fluorescence intensities are similar throughout the organism. Additionally, the dynamic range of the data is limited (see the following Processing details) and the changes in fluorescence during development are relatively small. The initial high Shannon entropy of the raw data in Table \ref{tab:metrics_fm} is due to its high background noise, which is reduced after smoothing, as indicated by the sharp drop in the entropy while the standard deviation shows relative consistency. Then, the use of MCLAHE increases the entropy again, together with the large changes in other metrics, this time due to the contrast enhancement. Similarly to the previous example, the 4D CLAHE outperforms its 3D counterpart overall because of the lower MSE or, equivalently, the higher PSNR of the 4D results, indicating a higher similarity to the raw data. In other contrast metrics such as the standard deviation and the Shannon entropy, the 4D and 3D results have very close values, indicating the complexity of judging image contrast by a single metric. Visualization of the dynamics in Fig. \ref{fig:embryo}(e)-(l) also shows that the 4D CLAHE-processed data preserve more of the fluorescence intensity change than its 3D counterpart, while maintaining a high cell-to-background contrast. More complete comparisons of unprocessed and processed data are presented in the supplementary videos 5-6. The contrast-enhanced embryo development data potentially allow better tracking of cellular lineage and dynamics \cite{Kretzschmar2012,Amat2014}, which are challenging due to sparse fluorescent labeling.
\begin{table}[htb!]
\caption{Contrast metrics for fluorescence microscopy data}
\setlength{\tabcolsep}{3pt}
\centering
\begin{tabular}{|p{55pt}|p{60pt}|p{30pt}|p{30pt}|p{30pt}|}
\hline
Dataset & MSE($\times 10^{-3}$) & PSNR & STD & ENT \\
\hline
Raw & - & - & 0.0316 & 1.1284 \\
Smoothed & 0.300 & 173.7 & 0.0265 & 0.5141 \\
3D CLAHE & 7.767 & 159.6 & 0.1048 & 0.5262 \\
4D CLAHE & 5.667 & 160.9 & 0.0921 & 0.5241 \\
\hline
\end{tabular}
\label{tab:metrics_fm}
\end{table}

\noindent\textbf{Processing details.} The raw 4D fluorescence microscopy data have a size of 512$\times$512$\times$109$\times$144 in the ($x$, $y$, $z$, $t$) dimensions. They were first denoised by a median filter with a kernel size of (2, 2, 2, 1). In the application of 4D CLAHE, the kernel size of choice was (20, 20, 10, 25) and the same kernel size in the first three dimensions were used for 3D CLAHE to enable direct comparison. For both MCLAHE procedures, the clip limit was set at 0.25 and the number of histogram bins at 256. The intensities in the raw data were given as 8 bit unsigned integers resulting in only 256 possible values. Hence, only the GHR setting was used because the AHR setting would result in bins smaller than the resolution of the data. Processing with MCLAHE ran on the same server as for the photoemission spectroscopy data (see Section IV.A). The total runtime for processing the whole dataset using only CPUs was about 26 mins. Similar to the photoemission case study, the speed-up by GPU usage was tracked. The total runtime for processing the first 8 time frames of the dataset was 32 s on the GPU versus 85 s only on CPUs, representing a 2.7-fold speed-up.

\section{Perspectives}
While we have presented the applications of the MCLAHE algorithm to real-world datasets of up to multiple gigabytes in size, its current major performance limitation is in the memory usage, since the data needs to be loaded entirely into the RAM (of CPUs or GPUs), which may be challenging for very large imaging and spectroscopy datasets on the multi-terabyte scale that are becoming widely available \cite{Ouyang2017}. Future improvements on the algorithm implementation may include distributed handling of chunked datasets to enable operation on limited hardware resource by loading each time only a subset of the data. In addition, the number of nearest-neighbor kernels currently required for intensity mapping interpolation increases exponentially with the dimensionality $D$ of the data (see Section III.C). For datasets with $D < 10$, this may not pose an outstanding issue, but for even higher-dimensional datasets, new strategies may be developed for approximate interpolation of selected neighboring kernels to alleviate the exponential scaling.

On the other hand, the applications of MCLAHE are not limited by the examples given in this work but are open to other types of data. It is especially beneficial to the preprocessing of high dimensional data with dense sampling produced by various fast volumetric spectroscopic and imaging techniques \cite{Flannigan2012,Lu2014,Jahr2015,Ozbek2018,Miller2014} for improving the performance of feature annotation and extraction tasks.

Furthermore, the call for extension of the image processing toolkits in 2D and 3D to higher-dimensional imaging datasets also motivates the dimensional extension of more recent CE procedures, such as those in \cite{Dabass2019,Mehrish2019,Tan2019}, which will provide a wider choice of algorithms for multidimensional image processing and comprehensive comparison of the algorithm performance on data with various characteristics.

\section{Conclusion}
We present a flexible and efficient generalization of the CLAHE algorithm to arbitrary dimensions for contrast enhancement of complex multidimensional imaging and spectroscopy datasets. Our algorithm, the multidimensional CLAHE, improves upon previous lower-dimensional equivalents \cite{Pizer1987,Zuiderveld1994,Amorim2018} by its unified treatment of image boundaries, flexible kernels size selection, adaptive histogram range determination. Its parallelized implementation in Tensorflow allows computational acceleration with multiple CPUs and GPUs. We demonstrate the effectiveness of multidimensional CLAHE by visual analysis and contrast quantification in case studies drawn from different measurement techniques, namely, 4D (3D+time) photoemission spectroscopy and 4D fluorescence microscopy, with the capabilities of producing densely sampled high dimensional data. In the example applications, our algorithm greatly improves and balances the visibility of multidimensional image features in diverse intensity ranges and neighborhood conditions. We further show that the best overall performance in each case comes from the simultaneous application of multidimensional CLAHE to all data dimensions, in line with the observation for applying CLAHE to 3D data \cite{Amorim2018}. In addition, we provide the implementation of multidimensional CLAHE in an open-source codebase to assist its reuse and integration into existing image analysis pipelines in various domains of natural sciences and engineering.

\section*{Acknowledgments}
We thank S. Sch\"{u}lke, G. Schnapka at the GNZ (Gemeinsames Netzwerkzentrum) in Berlin and M. Rampp at the MPCDF (Max Planck Computing and Data Facility) in Garching for computing supports. We thank S. Dong and S. Beaulieu for performing the photoemission spectroscopy measurement on a tungsten diselenide sample at the Fritz Haber Institute. V. Stimper thanks U. Gerland for administrative support. The work has received funding from the Max Planck Society, including BiGmax, the Max Planck Society's Research Network on Big-Data-Driven Materials Science, and funding from the European Research Council (ERC) under the European Union's Horizon 2020 research and innovation program (Grant Agreement No. ERC-2015-CoG-682843).

%\bibliographystyle{IEEEtran}
%\bibliography{literature}

\end{document}